\documentclass[11pt,preprint]{aastex}

\shorttitle{Evolution of Galaxy Luminosity Function}
\shortauthors{Ramos et al.}

\begin{document}

\title{Evolution of Galaxy Luminosity Function Using Photometric Redshifts}

\author{B. H. F. Ramos\altaffilmark{1,2}, P. S. Pellegrini\altaffilmark{1,2}, C. Benoist\altaffilmark{3,2},
L. N. da Costa\altaffilmark{1,2}, M. A. G. Maia\altaffilmark{1,2},
M. Makler\altaffilmark{4,2}, R. L. C. Ogando\altaffilmark{1,2}, F.
de Simoni\altaffilmark{1,2}, A. A. Mesquita\altaffilmark{5} }

\altaffiltext{1}{Observat\'orio Nacional, Rua Gal. Jos\'e Cristino
77, Rio de Janeiro, RJ - 20921-400, Brazil; ramos@linea.gov.br,
pssp@linea.gov.br, ldacosta@linea.gov.br, maia@linea.gov.br,
ogando@linea.gov.br, fsimoni@linea.gov.br}
\altaffiltext{2}{Laborat\'orio Interinstitucional de e-Astronomia
- LIneA, Rua Gal. Jos\'e Cristino 77, Rio de Janeiro, RJ -
20921-400, Brazil} \altaffiltext{3}{Universit\'e de Nice
Sophia-Antipolis, CNRS, Observatoire de la C\^ote d'Azur, UMR 6202
CASSIOP\'EE, BP 4229, F-06304 Nice Cedex 4, France;
benoist@oca.eu} \altaffiltext{4}{Centro Brasileiro de Pesquisas
F\'\i sicas, Rua Dr. Xavier Sigaud, 150, Rio de Janeiro 22290-180
RJ, Brazil; martin@cbpf.br}
\altaffiltext{5}{Universidade Federal
do Rio de Janeiro, Observat\'orio do Valongo, Ladeira do Pedro
Antonio, 43, Rio de Janeiro 20080-090, RJ, Brazil;
albertoalves@on.br}

\begin{abstract}

We examine the impact of using photometric redshifts for studying
the evolution of both the global galaxy luminosity function (LF)
and that for different galaxy types. To this end we compare the
LFs obtained using photometric redshifts from the CFHT Legacy
Survey (CFHTLS) D1 field with those from the spectroscopic survey
VIMOS VLT Deep Survey (VVDS) comprising $\approx$4800 galaxies. We
find that for $z \leq 2.0$, in the interval of magnitudes
considered by this survey, the LFs obtained using photometric and
spectroscopic redshifts show a remarkable agreement. This good
agreement led us to use all of the four Deep fields of the CFHTLS
comprising $\approx$386000 galaxies to compute the LF of the
combined fields and estimate directly the error in the parameters
based on the field-to-field variation. We find that the
characteristic absolute magnitude $M^*$ of Schechter fits fades by
$\approx 0.7$ mag from $z \approx 1.8$ to $z \approx 0.3$, while
the characteristic density $\phi^*$ increases by a factor of
$\approx 4$ in the same redshift interval. We use the galaxy
classification provided by the template fitting program used to
compute photometric redshifts and split the sample into galaxy
types. We find that these Schechter parameters evolve differently
for each galaxy type, an indication that their evolution is a
combination of several effects: galaxy merging, star formation
quenching and mass assembly. All these results are compatible with
those  obtained by different spectroscopic surveys such as VVDS,
DEEP2 and zCosmos, which reinforces the fact that photometric
redshifts can be used to study galaxy evolution, at least for the
redshift bins adopted so far. This is of great interest since
future very large imaging surveys containing hundreds of millions
of galaxies will allow to obtain important precise measurements to
constrain the evolution of the LF and to explore the dependence of
this evolution on morphology and/or color helping constrain the
mechanisms of galaxy evolution.

\end{abstract}

\keywords{galaxies: evolution --- galaxies: luminosity function}

\section{Introduction}

The galaxy luminosity function, or number density of galaxies as a
function of their luminosity, is a fundamental property of the
galaxy distribution, as it provides information on how visible
matter is distributed among galaxies of various luminosities at a
given epoch. Therefore, its evolution can be used to constrain
models of galaxy evolution and structure formation
\citep{2003ApJ...599...38B}. In order to set tight constraints on
these models, one would ideally divide galaxies into a variety of
subgroups and derive independent luminosity functions that are
known to vary significantly with respect to several physical
parameters such as redshift, color, galaxy type, star formation
rate, etc. With the advent of very large and deep galaxy surveys,
this has become possible, leading to a large number of works
\citep[e.g.,][]{2007ApJ...665..265F,2009MNRAS.400..429C,2009A&A...508.1217Z}.
However, in order to benefit from large, homogeneous and complete
samples, in most cases one has to rely only on the photometric
information with at most a partial spectroscopic coverage.

While the local luminosity function has been extensively studied
due to numerous spectroscopic surveys carried out over the years
\citep{1994ApJ...428...43M,1997AJ....113..185M,1998ApJ...503..617M,1999MNRAS.308..459F,2001AJ....121.2358B,2002MNRAS.333..133M},
probing its evolution to high redshifts still represents a
challenge. One possible way of doing that is to take advantage of
the recent multi-band photometric surveys, such as the
Canada-France-Hawaii Telescope Legacy Survey (CFHTLS), and use the
photometric redshift technique to estimate distances. The purpose
of this paper is to show that this is in fact possible, at least
using the redshift bins here adopted, yielding reliable results.
This is demonstrated using the VIMOS VLT Deep Survey (VVDS) data
and one of the Deep fields of CFHTLS. Based on this finding we use
all CFHTLS Deep fields to compute a combined luminosity function
and estimate the cosmic variance on their fitted Schechter
parameters at different redshifts. Finally, we use the information
provided by the template fitting routine used to estimate the
photometric redshift to split the sample into galaxy types, and
compare with the few results available in the literature. This is
relevant to investigate how reliable this approach can be to
explore the data that will eventually become public for surveys
such as the Dark Energy Survey (DES) and Large Synoptic Survey
Telescope (LSST).

In Section \ref{lfmethod} of this paper, the details on the method
used  to compute the luminosity function are discussed. The data
used is briefly described in Section \ref{data_desc}. In Section
\ref{lf_spec_samp}, the results of the comparison of the
luminosity functions computed using spectroscopic and photometric
redshifts are presented. In this section we also show the
evolution of the luminosity function derived for the combined
sample comprising all galaxies and split into different galaxy
types, which are then compared with the results of other authors.
We present a summary of our conclusions in Section \ref{concl}.

Throughout this paper, the cosmology used is $\Omega_{m} = 0.3$,
$\Omega_{\lambda} = 0.7$ and results are expressed in terms of
$h=H_0/100$.

\section{Estimating the Luminosity Function}
\label{lfmethod}

The luminosity function $\phi(M)dM$, expressing the number of
galaxies with absolute magnitudes in the interval $M$ and $M+dM$ per
unit volume, is calculated in intervals of redshift using the
$1/V_{max}$ method \citep{schmidt68}. Thus a galaxy of absolute
magnitude $M$ at redshift $z$ contributes to the luminosity function
with weight $1/V_{max}$, where $V_{max}$ is the maximum volume
within which the galaxy apparent magnitude is found between the catalog
apparent magnitude limits (also bounded by the redshift bin limits).

The absolute magnitudes are calculated as:
\begin{equation}
M = m - 25 - 5log(D_L) - K
\end{equation}
\noindent where $m$ is the apparent magnitude of the galaxy; the
luminosity distance is given by $D_L = (1+z) r$, where $r$ is the
comoving distance in $h^{-1}Mpc$; $K$ is the $K$-correction for
the shift of the observed spectrum with relation to rest frame
reference band wavelength and width. In this work we deal only
with the $i^\prime$-band in the AB photometric system, so $M$
stands for $M_{i^\prime , AB}$ and no correction for
intrinsic dust reddening was performed.

\begin{figure}
\plotone{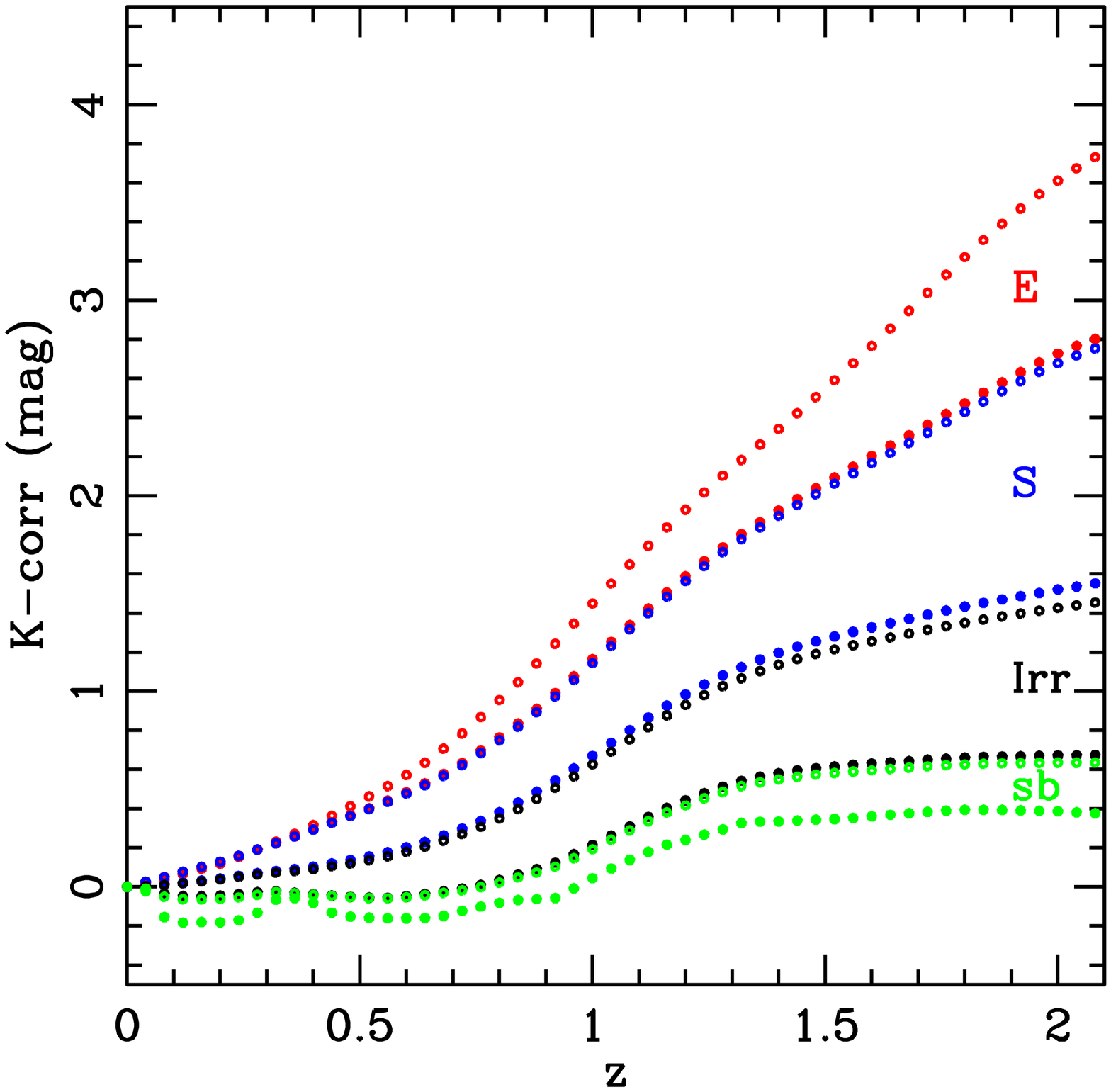} \caption{$K$-corrections for the $i^\prime$-band
for different galaxy types obtained from code Le Phare, where
labels indicate the range of corrections for the 62 spectral
types. \label{Kcorr}}
\end{figure}

Our source for $K$-corrections is the code Le Phare, which was run with
spectral energy distributions (SEDs) from \citet{1980ApJS...43..393C}
with the addition of 8 SEDs from \citet{1996ApJ...467...38K} \citep{2006A&A...457..841I}.
These $K$-corrections are shown in Figure \ref{Kcorr}, and
they are calculated for each galaxy based on the best fitting SED.

In the following analysis, we fit the usual \citet{1976ApJ...203..297S}
function:

\begin{eqnarray}
 \Phi(M)dM & = & 0.4\ln 10\phi^{*}10^{0.4(\alpha + 1)(M^{*} - M)}\times \nonumber \\
& & e^{-10^{0.4(M^{*} - M)}}dM
\end{eqnarray}

\noindent where $M^*$ is the characteristic absolute magnitude
and $\phi^*$ is the normalization factor. Fits for the Schechter
function were performed by a least square method using the
Marquardt-Levenberg algorithm with $M^*$, $\phi^*$ and $\alpha$ as
free parameters.

\section{Data}
\label{data_desc}

In this work, we used photometric data from the four Deep fields of the
CFHTLS from observations made with the MegaCam camera at the CFHT. Each
field (D1, D2, D3 and D4) covers $\approx 1$ square degree and was observed
in the $u*$, $g^\prime$, $r^\prime$, $i^\prime$, and $z^\prime$ bands
(Table \ref{infoCFHTLS} presents general information).

The CFHTLS-D1 field is particularly interesting for this work
since it has both photometric and spectroscopic observations from
VVDS. Photometry in the VVDS was carried out using $B$, $V$, $R$,
$I$ filters \citep{2006A&A...457..841I} and additional photometric
data in the $J$ and $K$ bands are available from
\citet{2005A&A...442..423I}. The spectra were obtained with the
VIMOS spectrograph at the VLT, for objects selected in the range
$17.5 \leq I_{AB} \leq 24.0$ \citep{2005A&A...439..845L}, yielding
a total of 6582 galaxies. We used only good quality spectroscopic
redshifts (quality flags 2, 3 and 4), to $z=2$, reducing the
sample to 4813.

In this analysis, we used the color catalogs processed by
Terapix\footnote{http://terapix.iap.fr/} for all four Deep
Fields (version T0003), but with a
different set of updated masks produced by one of the authors
(C.B.) to cover defective regions and those surrounding very
bright stars. This was performed in two steps: first by creating
automatic polygons centered on stars brighter than i=19.0 with a
shape including diffraction spikes and a size scaled to the star
magnitudes; second by correcting by hand these polygons in the
case of the brightest stars that show large ghosts and by adding
polygons to mask remaining defects such as satellite trails. From
these catalogs, we selected only objects classified as galaxies
by the Terapix group, from a morphological criterion based on
their location in a compactness (radius containing 50$\%$ of the
total flux) -magnitude distribution (see Terapix
site\footnotemark[\value{footnote}]), and
located outside masked regions, in order to avoid objects with
contaminated magnitudes. The photometric redshifts we use are
those provided in these catalogs, calculated by the Le Phare code,
which is based on a template fitting method and improved by an
empirical training set, which reduces the dispersion $\sigma_{\mid
\delta \mid z/(1+z)}$ from 0.047 to 0.029. Only galaxies with
available magnitudes in at least three bands had photometric
redshifts determined. The process to calculate such redshifts is
described in detail in \citet{2006A&A...457..841I} and was
conducted by the Terapix and VVDS teams.

We choose the $i^\prime$ band as the one defining the samples in
this work, since it is the closest to that ($I$-band) defining the
VVDS sample. Comparing $i^\prime _{AB}$ (CFHTLS) and $I_{AB}$
(VVDS) magnitudes down to $i^\prime_{AB}=24$ and $z<1.3$ we find
$i^\prime _{AB} = I_{AB} -0.11 \pm 0.16$. Due to this small
difference, we took for the $i^\prime$ band the same values for
the magnitude limit as the one in the $I$-band. Thus, whenever we
analyzed data from the CFHTLS individual fields (and specially to
compare the results with the VVDS survey) we used the limits $17.5
\leq i^\prime_{AB} \leq 24.0$, also restraining the analysis to $z
\leq 2.0$. When analyzing the sample from the combination of the
four CFHTLS areas, we extended these limits to $17.5 \leq
i^\prime_{AB} \leq 25.0$, corresponding to a 80$\%$ completeness
level \citep{2005A&A...439..863I}.

For a sample with magnitude limit of $i^\prime_{AB}=24$ (as in
this work), it is still possible to probe the faint end of the
luminosity function at $z \approx 1$ and determine a reliable
value for $\alpha$ when fitting the Schechter function. Due to the
lack of good sampling of faint objects at higher redshifts, we fix
the value of $\alpha$, usually for $z > 1.0$, to that obtained in
a previous bin where $\alpha$ is more reliably determined.

In the analysis of the following sections, we subdivide the
samples in individual CFHTLS fields in redshift intervals
[0.05-0.2], [0.2-0.4], [0.4-0.6], [0.6-0.8], [0.8-1.0], [1.0-1.3]
and [1.3-2.0] as used by \citet{2005A&A...439..863I}. When
analyzing the combined sample of four fields, the last intervals
considered are [1.3-1.6] and [1.6-2.0]. Table \ref{tab:appendix}
presents the number of galaxies in each redshift bin for each
sample analyzed in this work.

We discard galaxies with large i-band magnitude errors ($>$0.30
mag) and photometric redshift errors ($\epsilon _z > 0.5 z$). This
process discards 30$\%$ of the galaxies in the redshift bin
[0.05-0.2], less than 1$\%$ in the bin [0.2-0.4] and a negligible
amount in more distant bins. The total number of galaxies in the
four CFHTLS Deep fields resulting from all these selections
mentioned above is 385910 objects.

\section{Results}
\label{lf_spec_samp}

\subsection{Photometric versus Spectroscopic Redshifts }

In order to check how well photometric redshifts reproduce the
luminosity function obtained with spectroscopic redshifts, we
selected a sample of galaxies from the CFHTLS-D1 field in common
with the VVDS area, with both spectroscopic and photometric
redshift determinations. We call this data the Spectroscopic
Sample. We note that photometric redshifts for field D1 were
obtained in the CFHTLS from all available magnitudes which means
that for many galaxies of this Spectroscopic Sample typically 5 to
9 filters were used.

\begin{figure}
\plotone{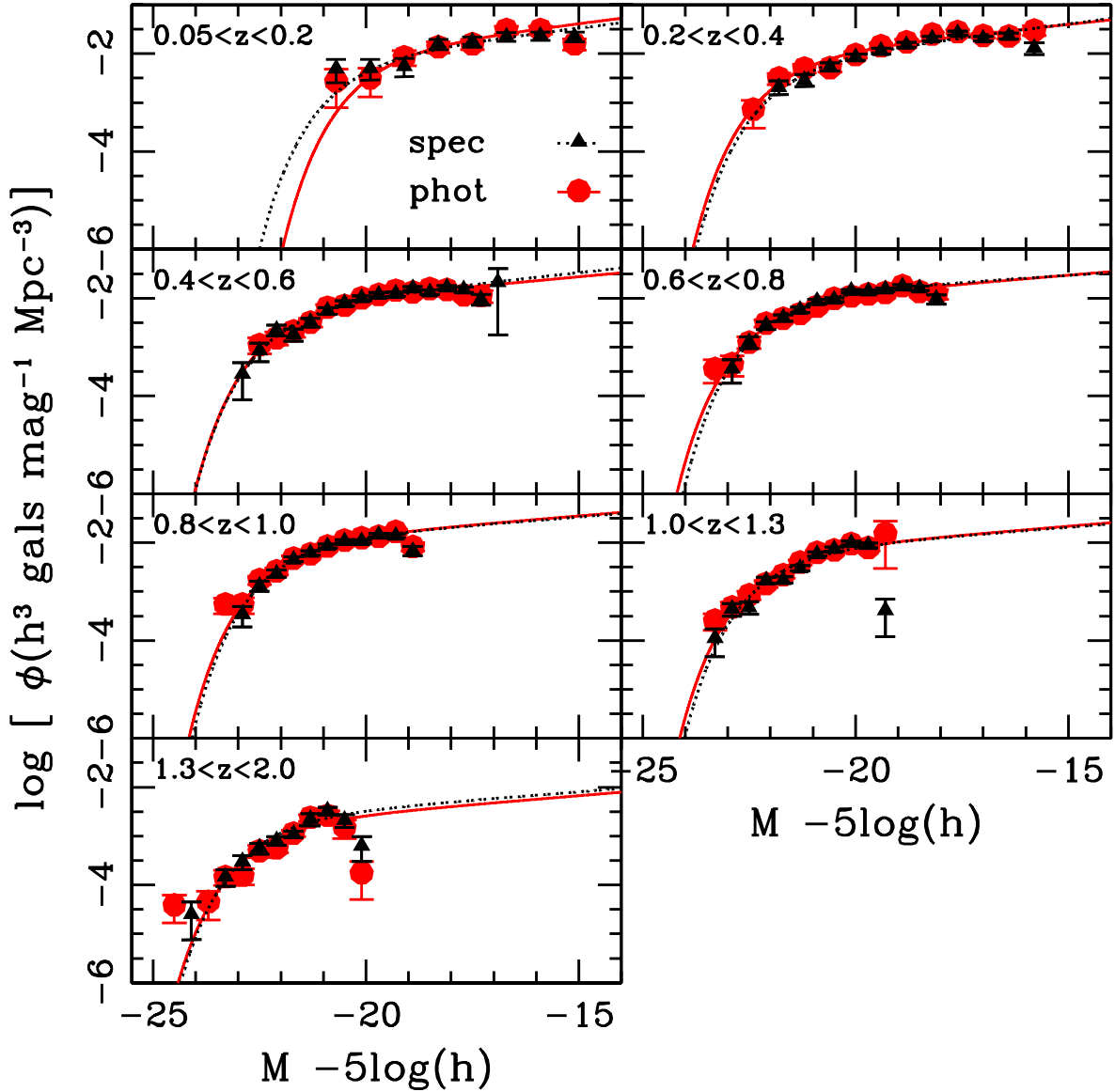} \caption{Luminosity functions in the
$i^\prime$-band for the Spectroscopic Sample in redshift bins.
Black triangles and dashed lines represent the best spectroscopic
data. Red dots and solid lines show the function determined with
photometric redshifts. \label{LF_spectsamp}}
\end{figure}

\begin{figure}
\plotone{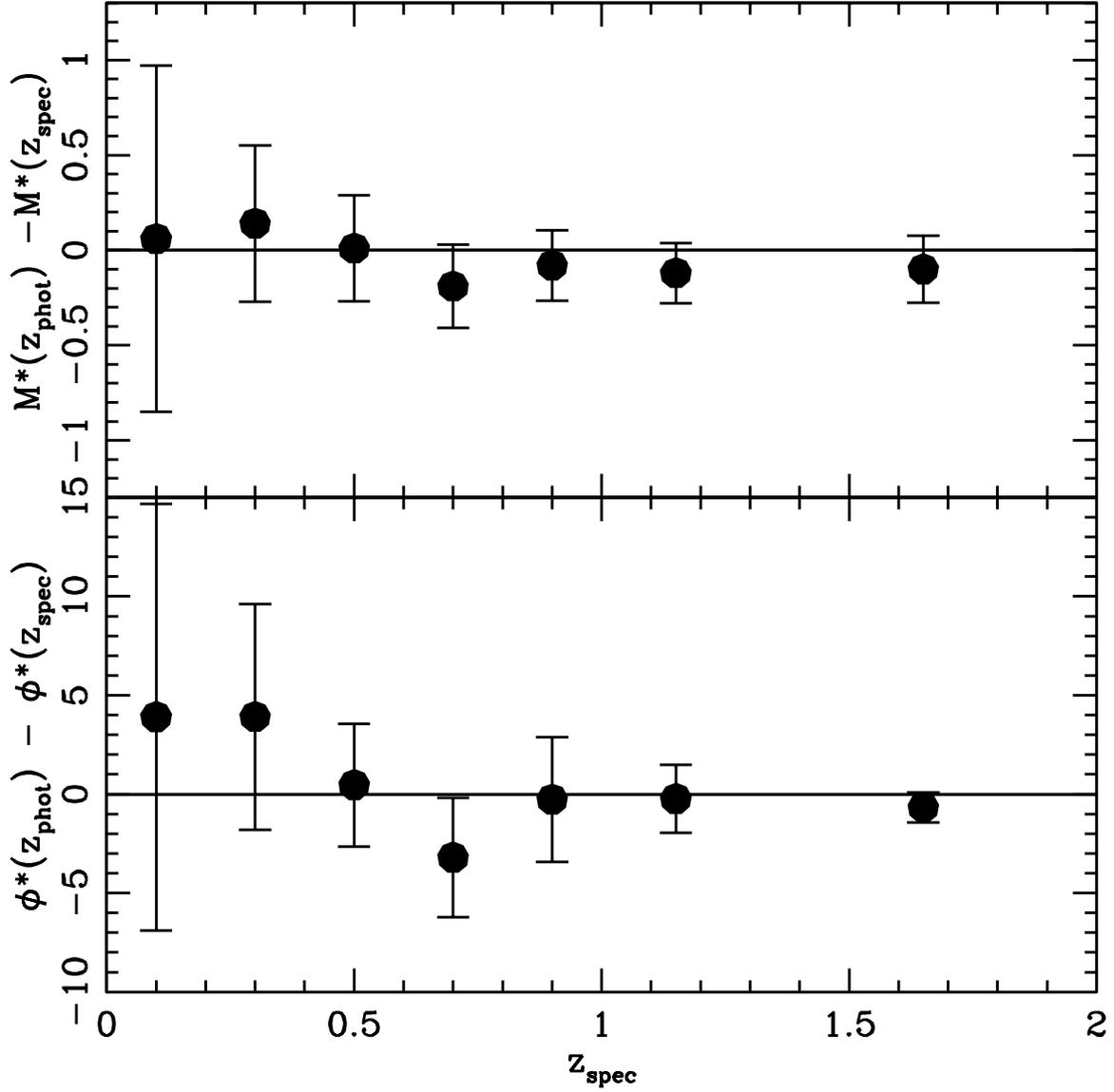} \caption{Comparison of $M^*$ (upper panel) and
$\phi^*$ (lower panel) for luminosity functions derived with
$z_{spec}$ and $z_{phot}$ data for the Spectroscopic Sample. Units
for the differences of $M^*$ and $\phi^*$ are mag and
10$^{-3}h^3$gals mag$^{-1}$Mpc$^{-3}$ respectively. \label{comp_spectsamp}}
\end{figure}

Luminosity functions were calculated in different redshift
intervals and Schechter fits were obtained. The derived $M^*$ and
$\phi^*$ are presented in Table \ref{tab:spectsamp} (results
obtained using both spectroscopic ($z_{spec}$) and photometric
($z_{phot}$) redshifts), while the luminosity functions and their
Schechter fits are shown in Figure \ref{LF_spectsamp}. We note
that these functions are not representative of a fair galaxy
sample since we do not apply the corrections for sampling biases
and redshift determination efficiency rate as described by
\citet{2005A&A...439..863I}. Nevertheless, since our primary
goal is to compare luminosity functions derived with $z_{phot}$
and $z_{spec}$, for the same set of objects, the fact that both
functions are not fair representations is not relevant and they
are computed only to evaluate their differences in $M^*$ and
$\phi^*$. These differences are shown in Figure
\ref{comp_spectsamp}.

At small redshifts ($0.05<z<0.2$) the uncertainties are large
and might reflect the small number of galaxies in this bin 
(see Table \ref{tab:appendix}) which prevents a reliable
determination of $M^*$ and $\phi^*$. At higher redshifts the
results show a remarkable agreement indicating that photometric
redshifts reproduce the spectroscopic results.

\subsection{Combining the CFHTLS Deep Fields}

We also calculate the luminosity function for all galaxies in the
VVDS area using photometric redshifts and compare the
characteristic parameters with those derived spectroscopically by
\citet{2005A&A...439..863I} yielding a good agreement between both
results. However, as shown below, the availability of four CFHTLS
areas similar in size can be used to improve the statistics
concerning this comparison as well as to investigate and quantify
the differences due to cosmic variance, at the $\approx 0.7$
(effective) square degree scale. Differently from field D1,
photometric redshifts for fields D2, D3 and D4 were obtained from
the set of filters $u*$, $g^\prime$, $r^\prime$, $i^\prime$, and
$z^\prime$. We estimate that for the combined four CFHTLS
fields, 76$\%$ of the objects had $z_{phot}$ calculated with at least
these five filters. Moreover as shown by \citet{2006A&A...457..841I},
the inclusion of the BVRIJK magnitudes for galaxies in the
CFHTLS-D1 field reduces the number of catastrophic events at
$z>1.3$ (due to the $J$ and $K$ bands) but reduces $\sigma_{\mid
\delta \mid z/(1+z)}$ from 0.029 to 0.026.
As discussed by \citet{2006A&A...457..841I} this is the larger
effect of including the additional VVDS filters and the J and K bands.
These authors also conclude that the accuracy of the photometric
redshifts decreases toward fainter apparent magnitudes and although
half of the objects with catastrophic errors are those classified as
starburst, the redshift accuracy is approximately independent of type.

\begin{figure}
\plotone{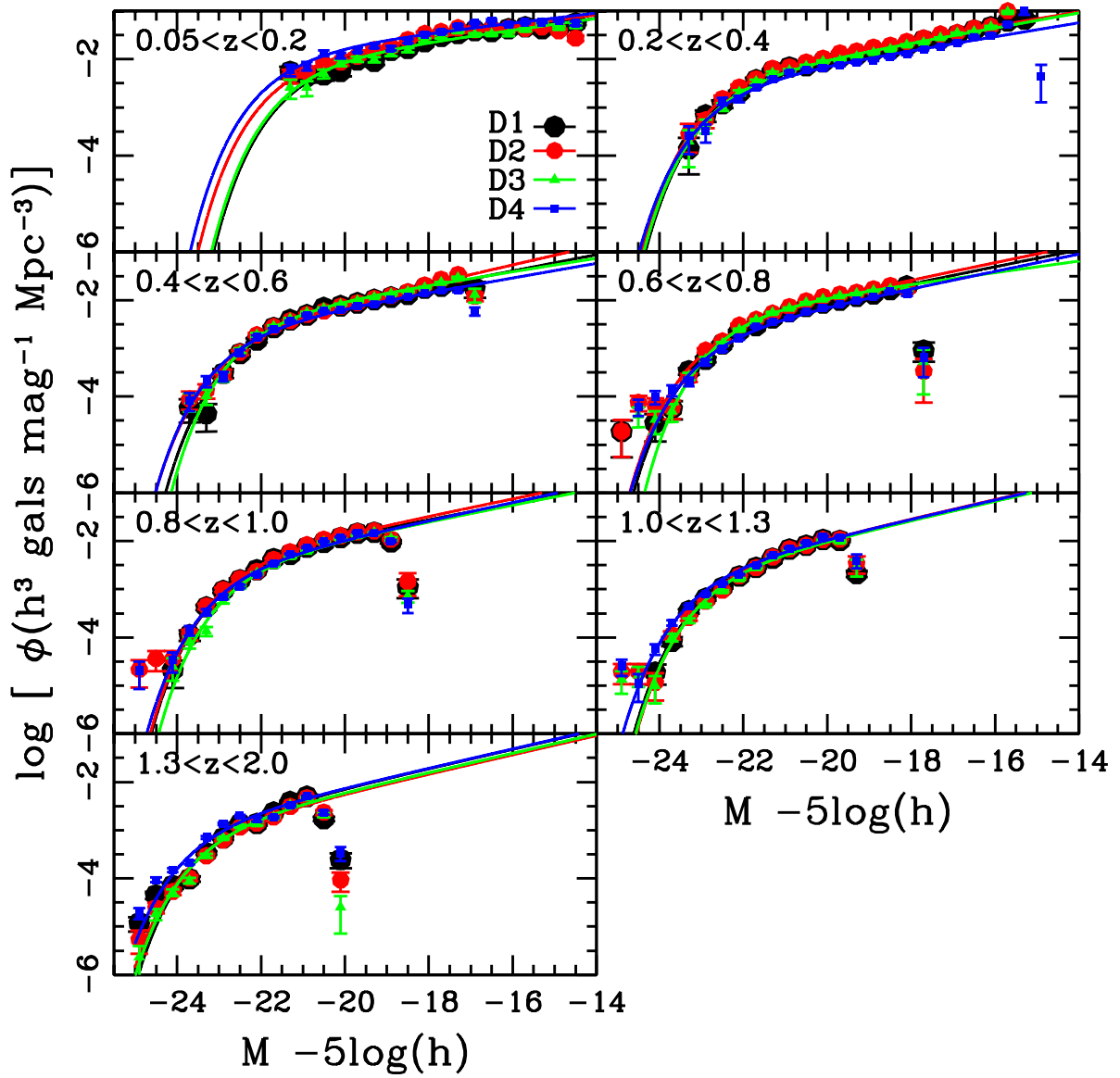} \caption{Luminosity functions in the
$i^\prime$-band for the four CFHTLS fields and their Schechter
fits. \label{LF_cfhtareas}}
\end{figure}

\begin{figure}
\plotone{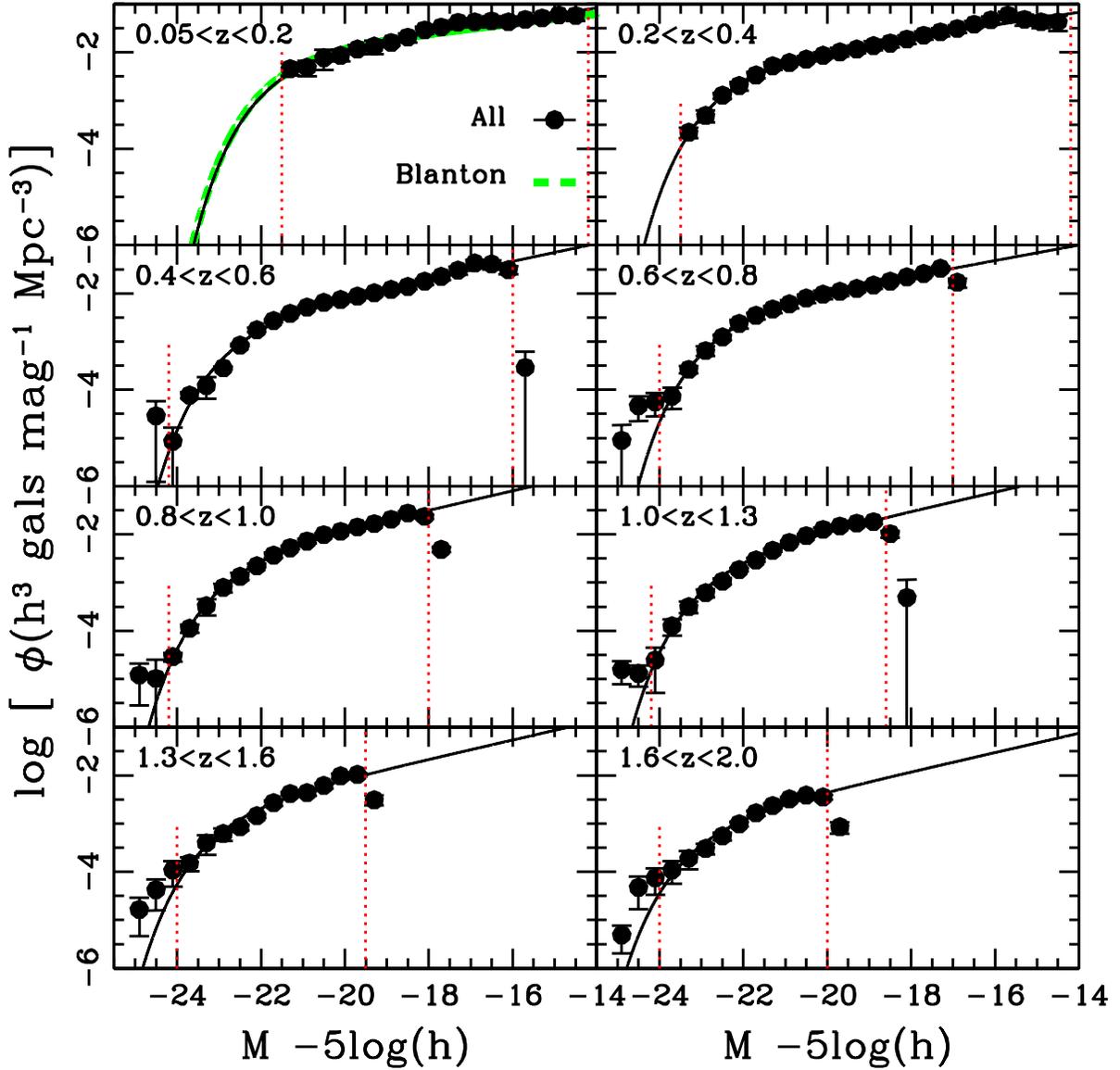} \caption{Luminosity functions in the
$i^\prime$-band and Schechter fits for the combined CFHTLS areas.
Red dotted vertical lines indicate the limits where the fits were performed.
The fit obtained by \citet{2001AJ....121.2358B} for the SDSS is
shown with a green dot-dashed line in the upper left panel.
\label{LF_cfhtareascomb}}
\end{figure}

The derived luminosity functions and respective Schechter fits for
each CFHTLS field are presented in Figure \ref{LF_cfhtareas} for
different redshift bins. Characteristic parameters $M^*$, $\phi^*$
and $\alpha$ of Schechter fits are presented in Table
\ref{tab:cfhtareas}. Although there is a general good agreement,
systematic differences are seen among the fields, showing that
cosmic variance is present in samples determined at the 0.7
deg$^{2}$ scale. In order to increase the number of objects in
each redshift bin and minimize cosmic variance, we merged the
samples of the four fields in a single combined CFHTLS sample.
The luminosity functions and their Schechter fits in each redshift bin,
as well as the magnitude range used for fitting,
are shown in Figure \ref{LF_cfhtareascomb}, while $M^*$, $\phi^*$
and $\alpha$ are presented in Table \ref{tab:cfhtareascomb}.

We note two features from Figure \ref{LF_cfhtareascomb}. One is
that with photometric redshifts it is possible to infer the
apparent non-linear shape of the faint end of the luminosity
functions seen at redshifts $z<0.6$. It is tempting to identify an
upturn of the function before the incompleteness cut in the
closest redshift bins. There is still a lot of discussion in the
literature concerning this issue \citep[e.g.,][]{2007ApJ...668..839R,
2008ApJ...672..198L,2009MNRAS.399.1106M,2009ApJ...692..778R,2010ApJ...721L..14B,
2010ApJ...725L.150O,2010ApJ...714L.202T} and it is beyond the
scope of this paper to address this question. We briefly mention
that some authors such as \citet{2005ApJ...631..208B} claim the
necessity to modify the Schechter function to correctly describe
the data, particularly the upturn of the function at faint
magnitudes. Anyway, it is becoming clear that this feature is a
consequence of a mix of galaxy populations
\citep[e.g.,][]{2009MNRAS.400..429C}, which at fainter magnitudes
are dominated by very late type galaxies. We mention this issue
again in the next section.

A second interesting feature is the systematic excess of galaxies
at the bright end ($M<-24$), a range which introduces significant
uncertainties on the fitted Schechter functions and which were
avoided in our fitting process. Indeed the small areas of the
individual CFHTLS fields may be subject to a particular larger
structure in a given redshift bin. For instance,
\citet{2007ApJS..172..254G} detected a galaxy cluster at $z=0.7$
in field D2 that could be the reason for the excess at the bright
end displayed by the respective luminosity function in Figure
\ref{LF_cfhtareas}. However, the combined luminosity functions in
Figure \ref{LF_cfhtareascomb} do show the excess for all
redshift bins where the bright end is observable. We note that
\citet{2009MNRAS.399.1106M} have also shown this excess in
their analysis of the SDSS-DR6 and claim this is the contribution
of galaxies with active nuclei, presenting a luminosity excess
compared to normal galaxies.

\begin{figure}
\epsscale{0.9}
\plotone{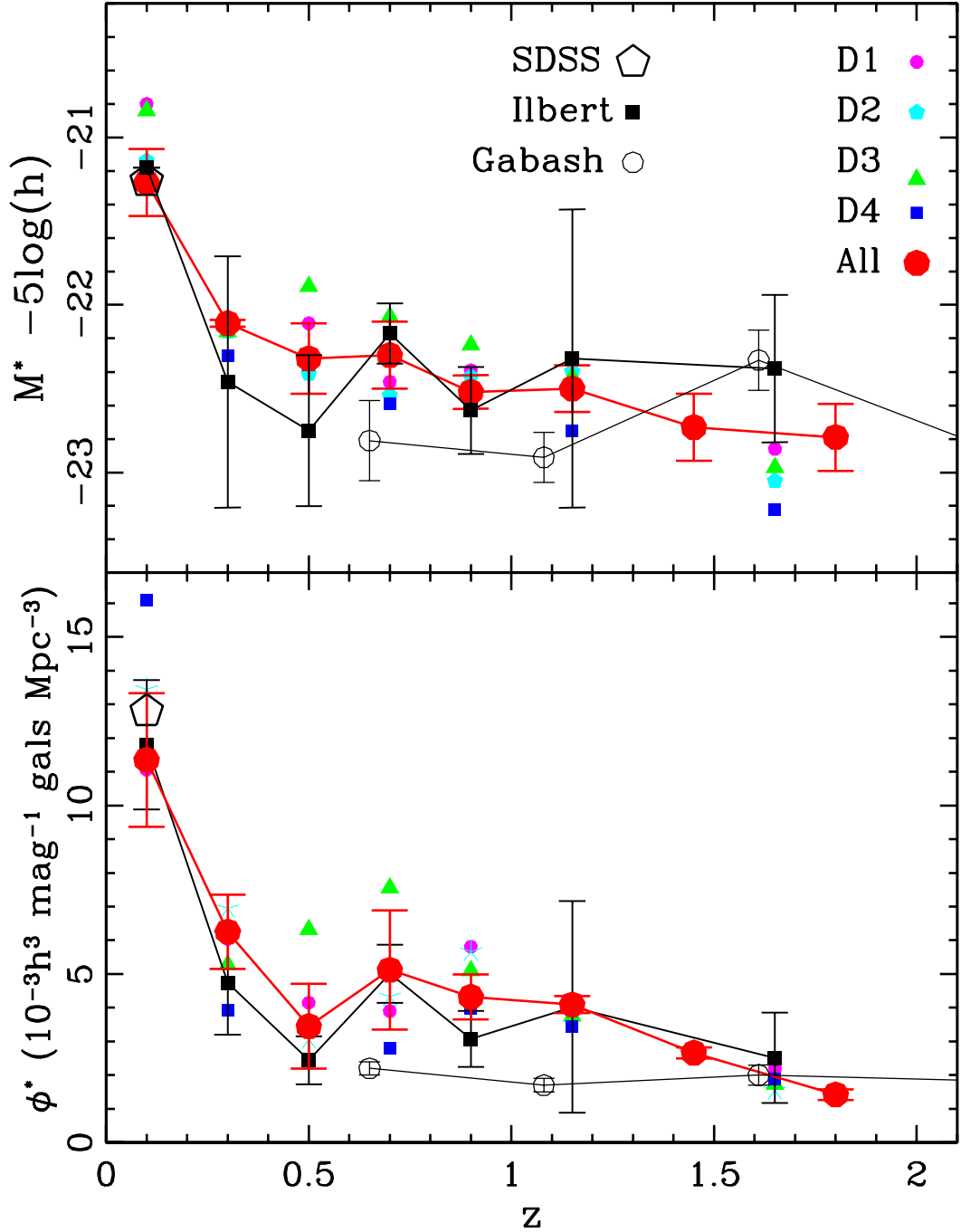} \caption{Evolution of $M^*$ and $\phi^*$ for the
combined CFHTLS areas (red dots). Error bars are the square root
of the variances from the four individual CFHTLS fields. The
results obtained by \citet{2005A&A...439..863I} are shown as black
squares, by \citet{2001AJ....121.2358B} for SDSS as a black open
pentagon and by \citet{2006A&A...448..101G} as black open circles.
\label{Mfievol_cfhtareascomb}}
\end{figure}

The evolution of $M^*$ and $\phi^*$ for the individual fields as
well as for the combined sample are shown in Figure
\ref{Mfievol_cfhtareascomb}. The results of the combined area
largely agrees with those of \citet{2005A&A...439..863I} for the
evolution of both $M^*$ and $\phi^*$, as well as with those for
the SDSS of \citet{2001AJ....121.2358B} for the DR1 and
\citet{2009MNRAS.399.1106M} for the DR6. These results indicate
for the global sample a slow fading  of $M^*$ with cosmic time by
$\approx 0.7$ mag from $z\approx 1.8$ to $z\approx 0.3$ and a much
faster dimming by $\approx 0.8$ mag to $z = 0$. Our results are
also compatible with those of \citet{2007ApJ...665..265F}, who
analyzing the DEEP2 and COMBO-17 surveys in the B-band, find a
dimming in $M^*$ of 1.2-1.3 mag from $z = 1$ to $z = 0$. Also,
\citet{2009A&A...508.1217Z} find that, in the zCOSMOS 10k sample,
$M^*$ fades in the B-band by $\approx 0.7$ mag from $z\approx 0.9$
to $z\approx 0.2$.

The lower panel of Figure \ref{Mfievol_cfhtareascomb} shows for
$\phi^*$ a similar trend as that of $M^*$ with an increase in
density by a factor of $\approx 4$ from $z\approx 1.8$ to
$z\approx 0.3$. The data suggest that there is also a significant
increase in the characteristic density of $M^*$ galaxies in the
$i^\prime$-band from $z = 0.3$ to $z = 0$, by a factor of $\approx
2$.  We also show in this figure the results of
\citet{2006A&A...448..101G} for the $i^\prime$ band, which uses
photometric redshifts and essentially indicates no evolution for
$M^*$ and $\phi^*$.

We note that the plateau seen in the lower panel of Figure
\ref{Mfievol_cfhtareascomb} in the interval $1.1 > z > 0.5$ is
consistent with the decrease of the merging rate shown in Figure 5
of \citet {2006ApJ...638..686C}. Also noticeable in Figure
\ref{Mfievol_cfhtareascomb} is that, near $z = 0$, $M^*$ dims
rapidly, while $\phi^*$ also rises significantly, with these
values at z=0 agreeing  with those observed in SDSS. This will be
addressed in the next section.

Since the errorbars plotted in this figure for the combined area
are the variances of $M^*$ and $\phi^*$, for the four fields,
Figure \ref{Mfievol_cfhtareascomb} and Table \ref{tab:cfhtareas}
allow us to estimate that the effect of cosmic variance over $M^*$
is $\approx 0.15$ mag for these fields of size $\approx 0.7$
square degree for $z<1$. Similarly, the effect of cosmic variance
over $\phi^*$ from these areas is $\approx 1.20 \times
10^{-3}h^3$gals mag$^{-1}$Mpc$^{-3}$.

\subsection{Analysis per Galaxy Types}

\begin{figure}
\plotone{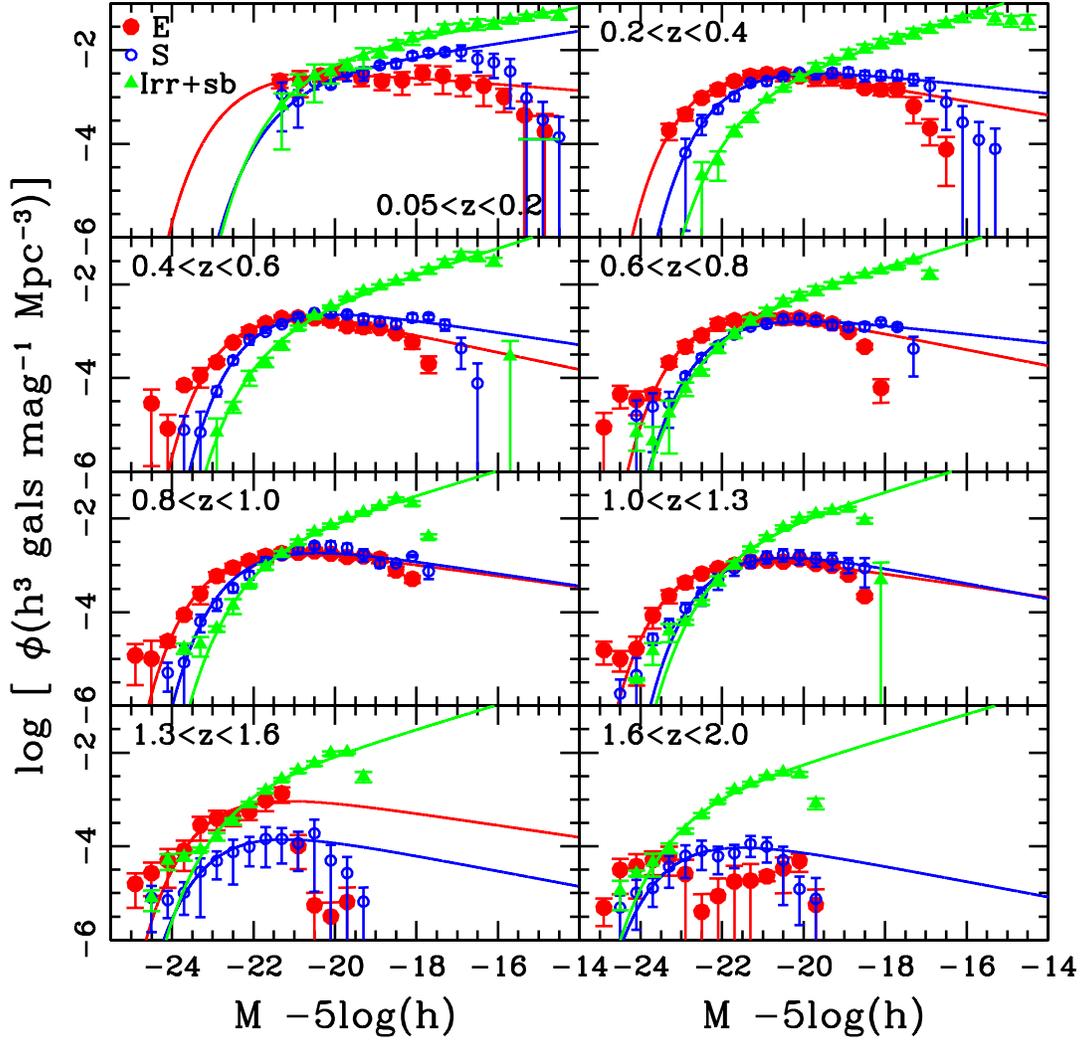} \caption{Luminosity functions in the
$i^\prime$-band and Schechter fits for the combined CFHTLS areas.
E-type galaxies are shown with red dots, S-type galaxies with blue
open circles and (Irr+sb)-type with green triangles.
\label{LF_cfhtareascombT}}
\end{figure}

As many authors have pointed out, the shape of the luminosity
functions largely depends on the galaxy population mix of the
sample under analysis \citep[e.g.,][]{1985AJ.....90.1759S,
2001AJ....121.2358B,2005MNRAS.356.1155C,2007ApJ...665..265F,
2009A&A...508.1217Z}. So, in order to investigate the relative
contributions of each galaxy type to the luminosity function evolution, we
have the advantage that the CFHTLS catalog includes the best fit
SED type that gave rise to the galaxy redshift. This is one of the
additional benefits of using a template fitting method to compute
photometric redshifts. We stress that although we refer to the
galaxy types in this work as E, S, Irr and starburst (sb), they should be
considered spectral types instead of morphological ones, since
they are attributed as a result of a SED fitting process.
Luminosity functions and respective Schechter fits for these types
are presented in Table \ref{tab:cfhttypes} and
shown in Figure \ref{LF_cfhtareascombT} in different redshift bins
for the combined CFHTLS areas. The combination of the
four fields was particularly important in the case of the samples separated
per galaxy type, in order for them to be statistically significant.

The galaxies classified here as (Irr+sb) outnumber the E and S
types for objects fainter than $M = -20$ at all redshifts and are
the dominant class for $z > 1.6$, as can be seen from Figure
\ref{LF_cfhtareascombT}. As noted by \citet{2006A&A...455..879Z},
these galaxies are responsible for most of the evolution (and
steepening) of the global luminosity function measured by
\citet{2005A&A...439..863I}. We also note the rise of the E class
at $z \approx 1.6$, forming even at these early epochs the
majority of the brightest objects ($M < -22$). As the
(Irr+sb)-type, S-type galaxies display in the CFHTLS data a
Schechter form at early epochs, in the range $1.6\leq z\leq 2.0$,
but present local deviations at the faint end at $z < 0.8$.
Moreover, Figure \ref{LF_cfhtareascombT} shows that, as reported
by different authors \citep[e.g.,][]{2005ApJ...631..208B,
2009MNRAS.400..429C}, the steepness of the global luminosity
function is due to the increasing number of (Irr+sb)-type
galaxies with decreasing luminosity. Also, the contribution of
the S-type galaxies may play a role in deviating the luminosity
function from a linear form at the faint end. We also note that
the excess at the bright end at $z>0.8$ is present in the latter
types as clearly seen in the (Irr+sb) luminosity functions.

\begin{figure}
\epsscale{0.9}
\plotone{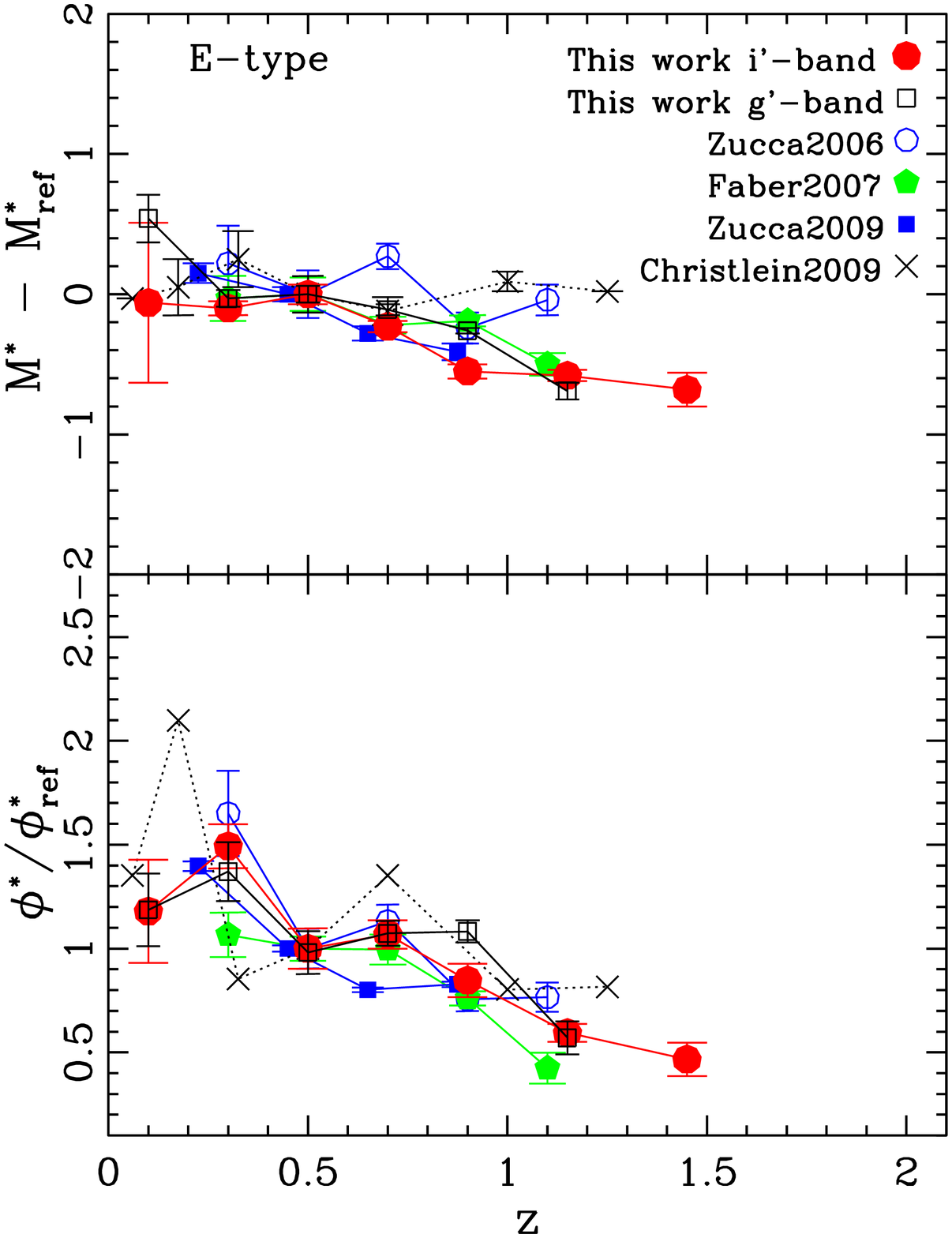} \caption{Evolution of $M^*$ (upper panel) and
$\phi^*$ (lower panel) for E-type galaxies of the combined CFHTLS
areas in the $i^{\prime}$-band (red dots) and in the
$g^{\prime}$-band (black empty squares). Results are shown with
reference to those at $z = 0.5$. Results from other surveys are
presented with symbols shown in the upper panel.
\label{Mfievol_cfhtareascombE}}
\end{figure}

\begin{figure}
\epsscale{0.9}
\plotone{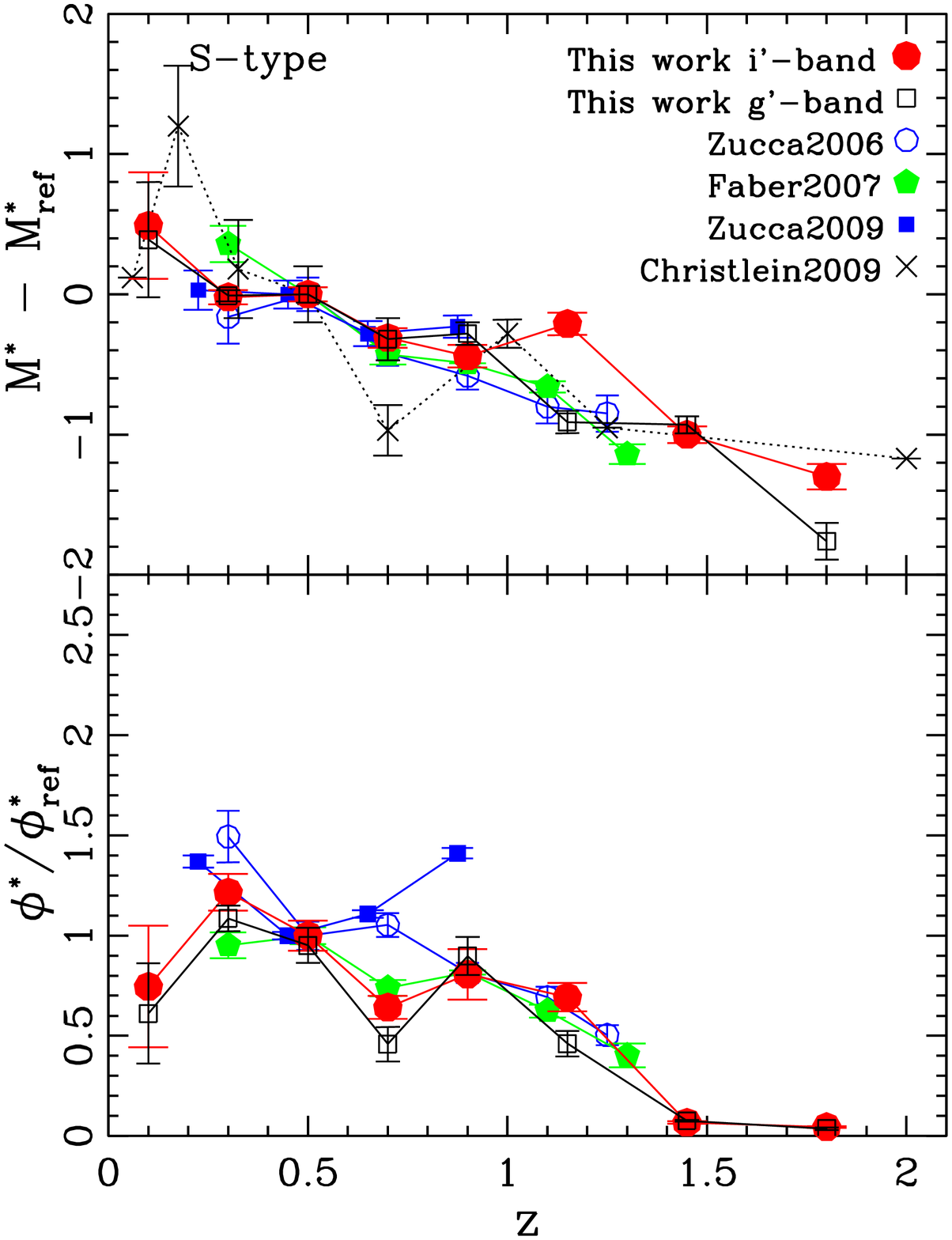} \caption{Evolution of $M^*$ (upper panel) and
$\phi^*$ (lower panel) for S-type galaxies of the combined CFHTLS
areas in the $i^{\prime}$-band (red dots) and in the
$g^{\prime}$-band (black empty squares). Results are shown with
reference to those at $z = 0.5$. Results from other surveys are
presented with symbols shown in the upper panel.
\label{Mfievol_cfhtareascombS}}
\end{figure}

\begin{figure}
\epsscale{0.9}
\plotone{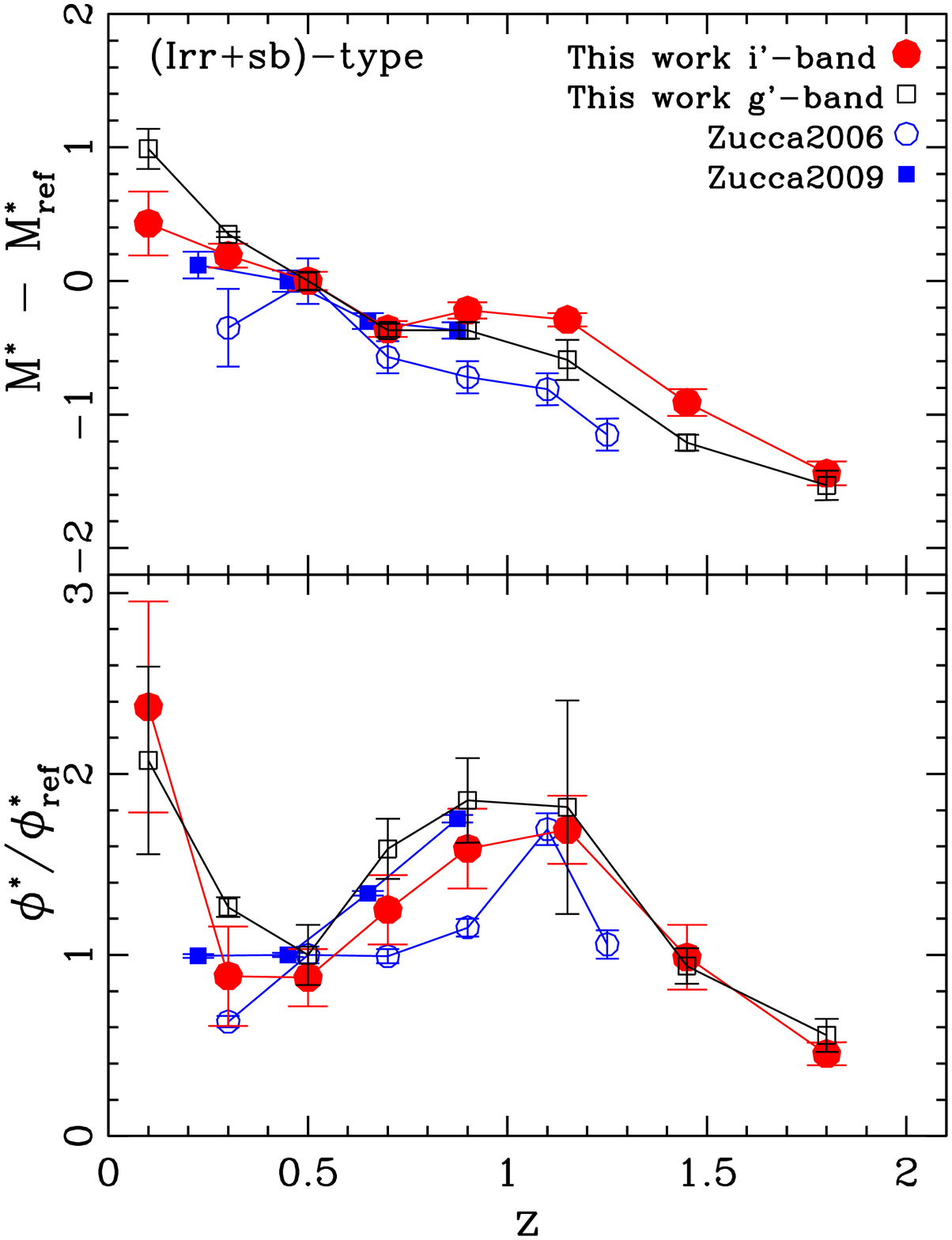} \caption{Evolution of $M^*$ (upper panel) and
$\phi^*$ (lower panel) for (Irr+sb)-type galaxies of the combined
CFHTLS areas in the $i^{\prime}$-band (red dots) and in the
$g^{\prime}$-band (black empty squares). Results are shown with
reference to those at $z = 0.5$. Results from other surveys are
presented with symbols shown in the upper panel.
\label{Mfievol_cfhtareascomb7}}
\end{figure}

In Figures \ref{Mfievol_cfhtareascombE},
\ref{Mfievol_cfhtareascombS} and \ref{Mfievol_cfhtareascomb7} we
show the evolution of $M^*$ and $\phi^*$ for the luminosity
functions of types E, S and (Irr+sb) galaxies. We present values
with reference to those at $z = 0.5$, termed here $M^*_{ref}$ and
$\phi^*_{ref}$, in order to compare with different authors. We
find that in the $i^\prime$-band:

\noindent - E-galaxies show a mild evolution of $M^*$, with a
dimming of $\approx 0.6$ mag from $z\approx 1.5$ to $z = 0.3$
while their number density increases in the same redshift interval
by a factor $\approx 3$. We have also considered evolutionary
effects using the $K$-corrections from \citet{2000Annis} derived
for the SDSS, and based on the evolutionary synthesis code
PEGASE-2 of \citet{1997A&A...326..950F}. Using the $K$-correction
representing passive evolution, we find that it does seem to
reproduce reasonably well the 0.6 mag dimming of this type of
galaxies, in the redshift range mentioned above;

\noindent - S-galaxies undergo a fading in their characteristic
value $M^*$ by 1.3 mag from $z\approx 1.8$ to $z = 0.3$ while
$\phi^*$ presents an increase by a factor of at least 2 from
$z\approx 1.3$ to $z\approx 0.3$;

\noindent - (Irr+sb)-galaxies shows a continuous decrease in
brightness $\approx 2$ mag in $M^*$ from $z\approx 1.8$ to $z =
0.3$. On the other hand $\phi^*$ presents a distinctive evolution:
it rises by a factor $\approx 4$ from $z \approx 1.8$ to $z
\approx 1.2$, then proceeds in the reverse sense with a decrease
in density by a factor $\approx 1.8$ to $z = 0.3$. We note
that this galaxy type presents lower accuracies of photometric
redshifts determinations as shown in figure 8 of
\citet{2006A&A...457..841I};

 \noindent - Inspection of table \ref{LF_cfhtareascombT}
indicates that the slope of the faint end does not change
significantly for each galaxy type.

As can be verified from figures \ref{Mfievol_cfhtareascombE},
\ref{Mfievol_cfhtareascombS} and \ref{Mfievol_cfhtareascomb7}
these results are, in general, in very good agreement with what
was found by other surveys such as the photometric MUSYC-ECDFS
\citep{2009MNRAS.400..429C} and COMBO-17
\citep{2007ApJ...665..265F}, as well as the spectroscopic VVDS
\citep{2006A&A...455..879Z}, DEEP-2
\citep{2007ApJ...665..265F} and z-COSMOS
\citep{2009A&A...508.1217Z}.  In order to perform these
comparisons we assumed that, in the MUSYC-ECDFS, early type
galaxies correspond to our E-galaxies, while their late type
objects (not shown because present large variations) correspond to
our S-galaxies. Concerning the comparison with the VVDS, we
assumed that their Type 1 galaxies correspond to our objects
classified as E-type, while their Types 2 and 3 correspond to our
S-types and their Type 4 corresponds to our (Irr+sb)-galaxies.
When comparing with the DEEP-2 and COMBO-17 we assumed that their
red galaxies represent our E-type galaxies and their blue sample
represents our S-galaxies. The comparison with the z-COSMOS
results was done assuming that their Type 1 corresponds to our
E-galaxies, their Type 2 corresponds to our S-galaxies and their
Types 3 and 4 correspond to our (Irr+sb) galaxies. Small
discrepancies may have origin in the different criteria used to
define the galaxy types in these works and in our analysis.

It is interesting to note that the lower panels of Figures
\ref{Mfievol_cfhtareascombE} and \ref{Mfievol_cfhtareascomb7} show
that, in the range $z \approx 1.0-0.5$, most of the decline
$\phi^*$ for the (Irr+sb)-type galaxies is compensated by a
comparable  increase of the E-type galaxies. This is consistent
with the transformation of blue cloud galaxies into red sequence
objects proposed by several previous authors
\citep[e.g.,][]{2006ApJ...651..120B, 2007ApJ...665..265F,
2008MNRAS.389..567C, 2009MNRAS.393.1127S}.

From table \ref{tab:cfhttypes} we find that for $z \leq 0.2$ the
rise of $\phi^*$ in the lower panel of Figure
\ref{Mfievol_cfhtareascomb} is due to the contribution of the S
and (Irr+sb) populations which is twice of that of the E-type
galaxies, and which cause a decrease in the mean value of $M^*$.
Even though this is a possible explanation, one should also be
aware that in this interval photometric redshift errors are
larger, in particular because the galaxy population is dominated
by star-forming galaxies, which may impact the results.

We note that the good match between our results in the
$i^{\prime}$-band and those of other works concerning the
evolutionary trends of $M^*$ and $\phi^*$ is present even in the
case of surveys in the $B$-band such as the DEEP-2 and z-COSMOS.
In order to verify if the fact that we are comparing different
bands introduces inconsistencies in this comparison (a younger
stellar content predominantly contributes to the $B$-band while an
older stellar content to the $i^{\prime}$-band) we calculate the
luminosity functions in a bluer filter such as the
$g^{\prime}$-band.  The resulting $M^*$ and $\phi^*$ are also
plotted in figures \ref{Mfievol_cfhtareascombE},
\ref{Mfievol_cfhtareascombS} and \ref{Mfievol_cfhtareascomb7} and
present a good agreement with the $i^{\prime}$-band,
indicating that the latter may be used to study the cosmic
evolution of galaxy populations. We note that, when deriving the
luminosity density evolution, \citet{2007A&A...472..403T} find
measurable differences between the I and B bands in the VVDS data.
Since, in our results, $M^*(z)$ is slightly steeper in the
$g^\prime$-band for latter types we checked for more subtle
differences between the $i^{\prime}$ and $g^{\prime}$ bands
performing linear fits to the $M^*$ evolution for these filters,
in the range $0.3<z<1.1$. The difference, in this redshift
interval, is only 0.05 mag for the E-type galaxies and increases
from $\approx 0.2$ mag for the S-type galaxies to $\approx 0.4$
mag for the (Irr+sb)-type galaxies. These results are consistent
with a stronger evolution of the luminosity in the bluer bands,
which probes star formation better, and is more intense in later
type galaxies.

\section{Conclusions}
\label{concl}

In this paper we have used the spectroscopic and photometric data
available from VVDS and CFHTLS surveys to determine how well can
we reproduce the evolution of luminosity function based on large
photometric samples. This is a necessary exercise considering the
large photometric surveys been planned for this decade.

Our main conclusions are:

 1- Using a sample extracted from the VVDS data containing
galaxies with both spectroscopic and photometric redshifts  we
obtain very similar luminosity functions, reproducing with
$z_{phot}$ the Schechter parameters $M^*$ and $\phi^*$ obtained
with $z_{spec}$, within the errorbars;

2- We also find that we can reproduce with the photometric data of
the CFHTLS the evolution of $M^*$ and $\phi^*$ of the
spectroscopic VVDS sample as obtained by
\citet{2005A&A...439..863I}. These results indicate from $z
\approx$ 1.8 to 0.3 a mild dimming in $M^*$ of $\approx 0.7$ mag
while $\phi^*$ increases by a factor of $\approx 4$.

3- From the combined CFHTLS sample we estimate the cosmic variance
in surveys areas $\approx 0.7$ square degrees to be of order 0.15
in $M^*$ and of order 25\% in $\phi^*$ in the range $z = 0.3 -
1.3$;

4- The faint end slope of the global luminosity function
varies from $\approx 1.5$ to $\approx 1.3$ from $z = 0.9$ to $z = 0.3$.

5- We used template fitting from the available Terapix photometric
catalogs of the CFHTLS to assign galaxy types and derive type
dependent luminosity functions. We find that we can reproduce with
the combined CFHTLS sample the evolution of the characteristic
parameters of the luminosity function of existing spectroscopic
surveys such as the VVDS \citep{2006A&A...455..879Z}, DEEP2
\citep{2007ApJ...665..265F} and zCosmos
\citep{2009A&A...508.1217Z}. Evolution of $M^*$, as a dimming with
cosmic time, is similar for all galaxy types, but less pronounced
for E-type galaxies. The characteristic densities $\phi^*$ of E
and S type galaxies evolve similarly as an increase towards low
redshifts, while very late types show a distinctive evolution with
a decrease in density from $z = 1.2$ to 0.3.

6- We find also that the variation of the faint end slope of
the global luminosity function is essentially due to the evolving
mixture of galaxy types with the increasing proportion of E and S
galaxy types with decreasing redshift.

There are issues in the present analysis that deserve further
investigation with larger and deeper samples. For instance, the
(Irr+sb)-type galaxies, which seem to play an important role at $z
> 1.6$, are known to present large photometric redshift errors. A
more detailed  analysis involving the comparison with
spectroscopic redshifts would be of interest to evaluate the
reliability of the results obtained here.

Also, redefining the late-type sample, for instance by combining
S+Irr as a single class and starburst galaxies as a separate
class, may contribute to a better interpretation of the effects
involved in the evolution of the luminosity function and thus a
better understanding of the processes driving galaxy evolution.

Finally, re-computing the photometric redshifts without the
$u*$-band may provide limits to the redshift interval which can be
used to study the evolution of the luminosity function by future
surveys such as DES that will not include this filter.  Indeed
\citet{2006A&A...457..841I} show the importance of the $u*$-band
to photometric redshifts determinations in the ranges
$z_{phot}<0.4$ and $z_{phot}>3$.

The results of this paper show the ability of photometric
redshifts in estimating distances when a large database is used,
as will be the case of DES and future projects in the Petabyte
scale as LSST. Modern computational tools designed to treat this
kind of data provide powerful analyzes taking advantage of
multi-band photometry. Larger sky areas to be surveyed by these
projects will yield deeper insight concerning the characterization
of the galaxy luminosity function and its evolution.

\acknowledgments

Based on observations obtained with MegaPrime/MegaCam, a joint
project of CFHT and CEA/DAPNIA, at the Canada-France-Hawaii
Telescope (CFHT) which is operated by the National Research
Council (NRC) of Canada, the Institut National des Science de
l'Univers of the Centre National de la Recherche Scientifique
(CNRS) of France, and the University of Hawaii. This work is based
in part on data products produced at TERAPIX and the Canadian
Astronomy Data Centre as part of the Canada-France-Hawaii
Telescope Legacy Survey, a collaborative project of NRC and CNRS.
This paper makes use of photometric redshifts produced
jointly by Terapix and VVDS teams.

This research was carried out with the support of the
Laborat\'orio Interinstitucional de e-Astronomia (LIneA) operated
jointly by the Centro Brasileiro de Pesquisas F\'{\i}sicas (CBPF),
the Laborat\'orio Nacional de Computa\c{c}\~ao Cient\'{\i}fica
(LNCC) and the Observat\'orio Nacional (ON) and funded by the
Minist\'erio de Ci\^encia e Tecnologia (MCT).

We thank an anonymous referee for useful comments that
improved the paper. B.H.F.R. acknowledges financial support from
the CNPq (DTI grant 381.358/09-7 associated with the PCI/MCT/ON
Program). C.B. acknowledges the funding from the PCI/MCT-CBPF
Program and the hospitality of ICRA-CBPF and ON, where part of
this work was done. L.N.dC. acknowledges CNPq grants 476277 / 2006
and 304.202/2008-8, FAPERJ grants E-26/102.358/2009 and
E-26/110.564/2010, and FINEP grants 01.06.0383.00 and
01.09.0298.00. A.A.M. acknowledges a fellowship from DS-CAPES
Program.

\clearpage

\clearpage

\clearpage

\clearpage

\clearpage

\clearpage

\clearpage

\clearpage

\clearpage

\clearpage

\clearpage

\begin{deluxetable}{ccccccccc}
\tabletypesize{\scriptsize}
\tablecaption{Properties of CFHTLS Deep fields. \label{infoCFHTLS}}
\tablewidth{0pt}
\tablehead{ \colhead{} & \colhead{RA} & \colhead{DEC} & \colhead{} & \colhead{} & \colhead{} & \colhead{} & \colhead{} & \colhead{Eff.Area} \\
\colhead{Field} & \colhead{(J2000)} & \colhead{(J2000)} & \colhead{$u*$} & \colhead{$g^\prime$} & \colhead{$r^\prime$} & \colhead{$i^\prime$} & \colhead{$z^\prime$} &  \colhead{($\deg^{2}$)} \\
\colhead{(1)} & \colhead{(2)} & \colhead{(3)} & \colhead{(4)} & \colhead{(5)} & \colhead{(6)} & \colhead{(7)} & \colhead{(8)} & \colhead{(9)} }
\startdata
  D1 &  02:25:59 &  -04:29:40 &  26.5  &  26.4  &  26.1  &  25.9(25.1)  &  25.0 &  0.71  \\
  D2 &  10:00:28 &  +02:12:30 &  26.1  &  26.2  &  26.0  &  25.7(24.9)  &  24.9 &  0.69  \\
  D3 &  14:19:27 &  +52:40:56 &  25.9  &  26.6  &  26.4  &  26.2(25.7)  &  25.1 &  0.80  \\
  D4 &  22:15:31 &  -17:43:56 &  26.5  &  26.3  &  26.4  &  26.0(25.3)  &  25.0 &  0.74  \\
\enddata
\tablecomments{Col. (1): Field name. Col. (2) and (3): Central position of the field (RA presented in hours, minutes and seconds, and
DEC in degrees, arc minutes and arc seconds). Col. (4)-(8): Completeness magnitude (50$\%$, AB) of each band (for the $i\prime$ band we also show the 80$\%$ completeness magnitude inside parenthesis). Col. (9): Effective area (which exclude masked regions) in square degrees.}
 \end{deluxetable}

\clearpage

\begin{deluxetable}{ccccccccccc}
\tabletypesize{\scriptsize}
\tablecaption{Galaxy distribution with redshift. \label{tab:appendix}}
\tablewidth{0pt}
\tablehead{\colhead{$\Delta z$} & \colhead{SS} &
 \colhead {D1} & \colhead{D2} &
\colhead{D3} & \colhead{D4}  & \colhead{E} & \colhead {S} & \colhead{Irr+sb} & \colhead{Tot} \\
\colhead{(1)} & \colhead{(2)} & \colhead{(3)} & \colhead{(4)} &
\colhead{(5)} & \colhead{(6)} & \colhead{(7)} & \colhead{(8)} &
\colhead{(9)} & \colhead{(10)}   }
\startdata
0.05-0.2 &201 &2346  &2481  &2549 &3390  &722   &1569 &11174 &13465\\
0.2-0.4  &721 &7327  &8616  &8452 &5515  &3762  &4698 &36903 &45363\\
0.4-0.6  &784 &7345  &7934  &9242 &6541  &4633  &6636 &44617 &55886\\
0.6-0.8 &1043 &8483 &10364 &11218 &7320  &7501  &7374 &48798 &63673\\
0.8-1.0 &1082 &9314  &9875  &9534 &9584  &9096 &10907 &51569 &71572\\
1.0-1.3  &708 &7497  &7541  &8240 &9508  &8598 &11568 &55710 &75876\\
1.3-2.0  &274 &5022  &4200  &5073 &5470 & \nodata & \nodata & \nodata & \nodata \\
1.3-1.6  & \nodata & \nodata & \nodata & \nodata & \nodata  &3130 &1046 &31928 &36104\\
1.6-2.0  & \nodata & \nodata & \nodata & \nodata & \nodata  &453 &924 &22594 &23971\\
\enddata
\tablecomments{Col. (1): Redshift bin. Col. (2): Spectroscopic
Sample. Col. (3)-(6): CFHTLS fields. Col. (7)-(9): Subsamples of
galaxy types. Col. (10): Combined data of the four fields. Columns
(2)-(6) refer to samples with limits $17.5 \le i'_{AB}\le 24$ and
$z_{phot}\le 2.0$. Columns (7)-(10) refer to samples with limits
$17.5 \le i'_{AB}\le 25$ and $z_{phot}\le 2.5$.}
\end{deluxetable}

\clearpage

\begin{deluxetable}{cccc}
\tabletypesize{\scriptsize}
\tablecaption{Schechter function parameters for the Spectroscopic Sample. \label{tab:spectsamp}}
\tablewidth{0pt}
\tablehead{
\colhead{} & \colhead{$M^*$} & \colhead {$\phi^*$} & \colhead{}  \\
\colhead{$<z>$} & \colhead{(mag-5log($h$))} & \colhead {(10$^{-3}h^3$gals mag$^{-1}$Mpc$^{-3}$)} & \colhead{$\alpha$}  \\
\colhead{(1)} & \colhead{(2)} & \colhead{(3)} & \colhead{(4)}  }
\startdata
\multicolumn{4}{c}{$z_{spec}$} \\
\hline
0.10 &-20.17 $\pm$ 1.10 &10.07 $\pm$ 9.57 &-1.27 $\pm$ 0.21 \\
0.30 &-21.44 $\pm$ 0.31 & 8.40 $\pm$ 2.74 &-1.28 $\pm$ 0.07 \\
0.50 &-21.71 $\pm$ 0.14 & 8.22 $\pm$ 1.75 &-1.24 $\pm$ 0.07 \\
0.70 &-21.65 $\pm$ 0.11 &13.53 $\pm$ 1.91 &-1.14 $\pm$ 0.06 \\
0.90 &-21.70 $\pm$ 0.15 &12.74 $\pm$ 2.44 &-1.17 $\pm$ 0.10 \\
1.15 &-21.69 $\pm$ 0.13 & 7.87 $\pm$ 1.48 &-1.17            \\
1.65 &-22.16 $\pm$ 0.12 & 2.85 $\pm$ 0.53 &-1.17            \\
\hline
\multicolumn{4}{c}{$z_{phot}$} \\
\hline
0.10 &-19.59 $\pm$ 0.37 &14.40 $\pm$ 6.11 &-1.27 $\pm$ 0.12 \\
0.30 &-21.47 $\pm$ 0.20 &10.98 $\pm$ 2.98 &-1.23 $\pm$ 0.07 \\
0.50 &-21.70 $\pm$ 0.24 & 8.68 $\pm$ 2.56 &-1.20 $\pm$ 0.10 \\
0.70 &-21.84 $\pm$ 0.19 &10.33 $\pm$ 2.34 &-1.18 $\pm$ 0.08 \\
0.90 &-21.78 $\pm$ 0.11 &12.47 $\pm$ 2.02 &-1.18 $\pm$ 0.09 \\
1.15 &-21.81 $\pm$ 0.09 & 7.64 $\pm$ 0.88 &-1.18            \\
1.65 &-22.26 $\pm$ 0.13 & 2.18 $\pm$ 0.54 &-1.18            \\
\enddata
\tablecomments{Col. (1): Central value of redshift interval. Col. (2): Characteristic absolute magnitude
and its Poisson error. Col. (3): Characteristic density and its Poisson error. Col. (4): Faint-end slope
and its Poisson error (for $z \ge 1.0$ it is fixed to -1.17 and -1.18 when using $z_{spec}$ and $z_{phot}$,
respectively).}
\end{deluxetable}

\clearpage

\begin{deluxetable}{c c c c}
\tabletypesize{\scriptsize}
\tablecaption{Schechter function parameters for the CFHTLS fields. \label{tab:cfhtareas}}
\tablewidth{0pt}
\tablehead{
\colhead{} & \colhead{$M^*$} & \colhead {$\phi^*$} & \colhead{} \\
\colhead{$<z>$} & \colhead{(mag-5log($h$))} & \colhead {(10$^{-3}h^3$gals mag$^{-1}$Mpc$^{-3}$)} & \colhead{$\alpha$} \\
\colhead{(1)} & \colhead{(2)} & \colhead{(3)} & \colhead{(4)}  }
\startdata
\multicolumn{4}{c}{D1} \\
\hline
0.10 &-20.80 $\pm$ 0.30 &11.06 $\pm$ 2.93 &-1.31 $\pm$ 0.05 \\
0.30 &-22.10 $\pm$ 0.07 & 6.04 $\pm$ 0.65 &-1.38 $\pm$ 0.02 \\
0.50 &-22.11 $\pm$ 0.14 & 4.15 $\pm$ 1.15 &-1.42 $\pm$ 0.05 \\
0.70 &-22.46 $\pm$ 0.07 & 3.90 $\pm$ 0.60 &-1.44 $\pm$ 0.03 \\
0.90 &-22.39 $\pm$ 0.03 & 5.81 $\pm$ 0.40 &-1.41 $\pm$ 0.08 \\
1.15 &-22.47 $\pm$ 0.03 & 3.87 $\pm$ 0.23 &-1.50    \\
1.65 &-22.86 $\pm$ 0.10 & 2.20 $\pm$ 0.38 &-1.50    \\

\hline
\multicolumn{4}{c}{D2} \\
\hline
0.10 &-21.14 $\pm$ 0.29  &13.45 $\pm$ 3.52 &-1.29 $\pm$ 0.05 \\
0.30 &-22.16 $\pm$ 0.04  & 6.95 $\pm$ 0.48 &-1.36 $\pm$ 0.02 \\
0.50 &-22.41 $\pm$ 0.13  & 3.06 $\pm$ 0.74 &-1.50 $\pm$ 0.08 \\
0.70 &-22.54 $\pm$ 0.15  & 4.33 $\pm$ 1.35 &-1.45 $\pm$ 0.16 \\
0.90 &-22.43 $\pm$ 0.08  & 5.67 $\pm$ 0.77 &-1.45 $\pm$ 0.04 \\
1.15 &-22.40 $\pm$ 0.04  & 4.14 $\pm$ 0.35 &-1.50    \\
1.65 &-23.05 $\pm$ 0.00  & 1.55 $\pm$ 0.25 &-1.50    \\

\hline
\multicolumn{4}{c}{D3} \\
\hline
0.10 &-20.84 $\pm$ 0.21 &11.50 $\pm$ 2.45 &-1.30 $\pm$ 0.04 \\
0.30 &-22.16 $\pm$ 0.06 & 5.28 $\pm$ 0.46 &-1.39 $\pm$ 0.02 \\
0.50 &-21.89 $\pm$ 0.05 & 6.32 $\pm$ 0.61 &-1.35 $\pm$ 0.04 \\
0.70 &-22.07 $\pm$ 0.03 & 7.55 $\pm$ 0.47 &-1.30 $\pm$ 0.03 \\
0.90 &-22.24 $\pm$ 0.13 & 5.10 $\pm$ 1.30 &-1.43 $\pm$ 0.11 \\
1.15 &-22.43 $\pm$ 0.07 & 3.76 $\pm$ 0.42 &-1.50    \\
1.65 &-22.97 $\pm$ 0.08 & 1.73 $\pm$ 0.28 &-1.50    \\

\hline
\multicolumn{4}{c}{D4} \\
\hline
0.10 &-21.29 $\pm$ 0.32  &16.08 $\pm$ 4.07 &-1.26 $\pm$ 0.04 \\
0.30 &-22.30 $\pm$ 0.11  & 3.94 $\pm$ 0.61 &-1.36 $\pm$ 0.03  \\
0.50 &-22.36 $\pm$ 0.10  & 3.38 $\pm$ 0.61 &-1.38 $\pm$ 0.04  \\
0.70 &-22.59 $\pm$ 0.11  & 2.79 $\pm$ 0.49 &-1.45 $\pm$ 0.05  \\
0.90 &-22.57 $\pm$ 0.12  & 3.98 $\pm$ 0.96 &-1.47 $\pm$ 0.08  \\
1.15 &-22.75 $\pm$ 0.04  & 3.45 $\pm$ 0.26 &-1.50     \\
1.65 &-23.22 $\pm$ 0.25  & 1.90 $\pm$ 0.52 &-1.50     \\
\enddata
\tablecomments{The meanings of columns are the same as in Table \ref{tab:spectsamp}. The value
of $\alpha$ for $z \ge 1.0$ is fixed to -1.50.}
\end{deluxetable}

\clearpage

\begin{deluxetable}{c c c c}
\tabletypesize{\scriptsize}
\tablecaption{Schechter function parameters for the combined CFHTLS fields. \label{tab:cfhtareascomb}}
\tablewidth{0pt}
\tablehead{
\colhead{} & \colhead{$M^*$} & \colhead {$\phi^*$} & \colhead{} \\
\colhead{$<z>$} & \colhead{(mag-5log($h$))} & \colhead {(10$^{-3}h^3$gals mag$^{-1}$Mpc$^{-3}$)} & \colhead{$\alpha$}  \\
\colhead{(1)} & \colhead{(2)} & \colhead{(3)} & \colhead{(4)} }
\startdata
0.10 &-21.28 $\pm$ 0.26 &11.24 $\pm$ 2.44 &-1.31 $\pm$ 0.03\\
0.30 &-22.11 $\pm$ 0.07 & 6.22 $\pm$ 0.64 &-1.34 $\pm$ 0.02 \\
0.50 &-22.31 $\pm$ 0.06 & 3.47 $\pm$ 0.49 &-1.46 $\pm$ 0.03 \\
0.70 &-22.30 $\pm$ 0.02 & 5.12 $\pm$ 0.22 &-1.40 $\pm$ 0.01 \\
0.90 &-22.52 $\pm$ 0.04 & 4.32 $\pm$ 0.39 &-1.50 $\pm$ 0.02 \\
1.15 &-22.48 $\pm$ 0.05 & 4.16 $\pm$ 0.32 &-1.50  \\
1.45 &-22.73 $\pm$ 0.09 & 2.63 $\pm$ 0.31 &-1.50  \\
1.80 &-22.76 $\pm$ 0.11 & 1.49 $\pm$ 0.23 &-1.50  \\
\enddata
\tablecomments{The meanings of columns are the same as in Table \ref{tab:spectsamp}. The value
of $\alpha$ for $z \ge 1.0$ is fixed to -1.50.}
\end{deluxetable}

\clearpage

\begin{deluxetable}{c c c c c c c}
\tabletypesize{\scriptsize}
\tablecaption{Schechter function parameters for the combined CFHTLS fields separated by galaxy type. \label{tab:cfhttypes}}
\tablewidth{0pt}
\tablehead{
\colhead{} & \colhead{$M^*$} & \colhead {$\phi^*$} & \colhead{} \\
\colhead{$<z>$} & \colhead{(mag-5log($h$))} & \colhead {(10$^{-3}h^3$gals mag$^{-1}$Mpc$^{-3}$)} & \colhead{$\alpha$} \\
\colhead{(1)} & \colhead{(2)} & \colhead{(3)} & \colhead{(4)}  }
\startdata
\multicolumn{4}{c}{E-type} \\
\hline
0.10 &-21.73 $\pm$ 0.57 &5.00 $\pm$ 1.05 &-0.83 $\pm$ 0.08 \\
0.30 &-21.77 $\pm$ 0.05 &6.33 $\pm$ 0.45 &-0.63 $\pm$ 0.04 \\
0.50 &-21.67 $\pm$ 0.07 &4.24 $\pm$ 0.41 &-0.54 $\pm$ 0.06 \\
0.70 &-21.90 $\pm$ 0.04 &4.53 $\pm$ 0.29 &-0.57 $\pm$ 0.04 \\
0.90 &-22.22 $\pm$ 0.05 &3.59 $\pm$ 0.34 &-0.70 $\pm$ 0.05 \\
1.15 &-22.25 $\pm$ 0.04 &2.52 $\pm$ 0.18 &-0.68 $\pm$ 0.04 \\
1.45 &-22.35 $\pm$ 0.12 &1.98 $\pm$ 0.34 &-0.68    \\
\hline
\multicolumn{4}{c}{S-type} \\
\hline
0.10 &-20.68 $\pm$ 0.38  &3.66 $\pm$ 1.49 &-1.33 $\pm$ 0.11 \\
0.30 &-21.19 $\pm$ 0.05  &5.97 $\pm$ 0.45 &-0.77 $\pm$ 0.03 \\
0.50 &-21.17 $\pm$ 0.05  &4.91 $\pm$ 0.37 &-0.67 $\pm$ 0.04 \\
0.70 &-21.48 $\pm$ 0.07  &3.15 $\pm$ 0.28 &-0.76 $\pm$ 0.04 \\
0.90 &-21.61 $\pm$ 0.08  &3.96 $\pm$ 0.62 &-0.67 $\pm$ 0.08 \\
1.15 &-21.38 $\pm$ 0.08  &3.40 $\pm$ 0.35 &-0.59 $\pm$ 0.07   \\
1.45 &-22.17 $\pm$ 0.06  &0.33 $\pm$ 0.03 &-0.59    \\
1.80 &-22.47 $\pm$ 0.09  &0.22 $\pm$ 0.02 &-0.59    \\
\hline
\multicolumn{4}{c}{(Irr+sb)-type} \\
\hline
0.10 &-20.55 $\pm$ 0.13 &7.58 $\pm$ 1.33 &-1.41 $\pm$ 0.03 \\
0.30 &-20.97 $\pm$ 0.07 &2.44 $\pm$ 0.36 &-1.70 $\pm$ 0.04 \\
0.50 &-21.16 $\pm$ 0.07 &2.40 $\pm$ 0.43 &-1.69 $\pm$ 0.05 \\
0.70 &-21.76 $\pm$ 0.07 &2.47 $\pm$ 0.36 &-1.67 $\pm$ 0.07 \\
0.90 &-21.45 $\pm$ 0.08 &4.34 $\pm$ 0.75 &-1.65 $\pm$ 0.06 \\
1.15 &-21.50 $\pm$ 0.12 &4.84 $\pm$ 1.17 &-1.66 $\pm$ 0.10 \\
1.45 &-22.09 $\pm$ 0.24 &2.85 $\pm$ 0.40 &-1.66    \\
1.80 &-22.55 $\pm$ 0.24 &1.42 $\pm$ 0.17 &-1.66    \\
\enddata
\tablecomments{The meanings of columns are the same as in Table \ref{tab:spectsamp}. The value
of $\alpha$ for $z \ge 1.3$ is fixed to that obtained in the previous redshift bin.}
\end{deluxetable}

\end{document}